\documentclass[aps, prx,twocolumn,superscriptaddress, reprint, nofootinbib, showpacs]{revtex4-1}

\usepackage{graphicx}
\usepackage{dcolumn}
\usepackage{dcolumn}
\usepackage{textgreek}
\usepackage{textcomp}
\usepackage{adjustbox}
\usepackage{multirow}

\def\KtCA{K$_2$Cr$_3$As$_3$}
\def\KoCA{KCr\textsubscript{3}As\textsubscript{3}}

\def\AoCA{\textit{A}Cr\textsubscript{3}As\textsubscript{3}}

\def\KHCA{KH$_x$Cr$_3$As$_3$}
\def\KHtfCA{KH$_{0.35}$Cr$_3$As$_3$}
\def\KHsnCA{KH$_{0.65}$Cr$_3$As$_3$}
\def\KHoCA{KHCr$_3$As$_3$}
\def\KtCA{K\textsubscript{2}Cr\textsubscript{3}As\textsubscript{3}}

\def\AtTA{$A_2TM_3$As$_3$}

\def\Tc{$T_c$}

\def\CAtube{Cr\textsubscript{6}As\textsubscript{6}}

\def\Psmt{$P\overline{6}m2$}

\def\Pstm{$P6_3/m$}

\begin{document}

\preprint{APS/123-QED}

\title{Tuning from Frustrated Magnetism to Superconductivity in Quasi-One-Dimensional \KoCA\ Through Hydrogen Doping}

\author{Keith M. Taddei}
\email[corresponding author ]{taddeikm@ornl.gov}
\affiliation{Neutron Scattering Division, Oak Ridge National Laboratory, Oak Ridge, TN 37831}
\author{Liurukara D. Sanjeewa}
\affiliation{Materials Science and Technology Division, Oak Ridge National Laboratory, Oak Ridge, TN 37831} 
\author{Bing-Hua Lei}
\affiliation{Department of Physics and Astronomy, University of Missouri, Missouri 65211, USA}
\author{Yuhao Fu}
\affiliation{Department of Physics and Astronomy, University of Missouri, Missouri 65211, USA}
\author{Qiang Zheng}
\affiliation{Materials Science and Technology Division, Oak Ridge National Laboratory, Oak Ridge, TN 37831}
\author{David J. Singh}
\affiliation{Department of Physics and Astronomy, University of Missouri, Missouri 65211, USA}
\author{Athena S. Sefat}
\affiliation{Materials Science and Technology Division, Oak Ridge National Laboratory, Oak Ridge, TN 37831}
\author{Clarina dela Cruz}
\affiliation{Neutron Scattering Division, Oak Ridge National Laboratory, Oak Ridge, TN 37831}

\date{\today}

\begin{abstract}

We report the charge doping of \KoCA\ via H intercalation. We show that the previously reported ethanol bath deintercalation of \KtCA\ to \KoCA\ has a secondary effect whereby H from the bath enters the quasi-one-dimensional \CAtube\ chains. Furthermore, we find that - contrary to previous interpretations - the difference between non-superconducting as-grown \KoCA\ samples and superconducting hydrothermally annealed samples is not a change in crystallinity but due to charge doping with the latter treatment increasing the H concentration in the CrAs tubes effectively electron-doping the 133 compound. These results suggest a new stoichiometry \KHCA , that superconductivity arises from a suppressed magnetic order via a tunable parameter and paves the way for the first charge-doped phase diagram in these materials. 
 
\end{abstract}


\maketitle

\section{\label{sec:intro}Introduction}

Superconductors are unique in the heirarchy of quantum materials. Arguably the start of the field, superconductivity (as manifested in unconventional superconductors) was one of the phenomenon which suggested topological classifications, gave rise to the Resonating Valence Bond theory now used to describe quantum spin-liquids and grew into the basis of many active research areas including Majorana Fermions and qubit design \cite{Bednorz1986,Anderson1987,Wen2017,NakamuraY1999,Read2000,Linder2015}. Yet unlike many of the topological quantum materials whose physics are theortically understood but which struggle to find physical realizations, unconventional superconductors exist, not infrequently, but have no settled upon microscopic theory leaving the search for new superconductors and the optimization of their properties unguided \cite{Wen2017, Fernandes2016,Lee2019,Balents2010,Norman2011} Instead, what has arisen is an empirical recipe born of a suprisingly universal phase diagram observed in unconventional superconductors where superconductivity is found near instabilities - namely of magnetic character \cite{Basov2011}.

It is therefore useful to find new unconventional superconductors which are structurally simple, have few orders (ideally none of which are coupled) and are easy to address theoretically (e.g. fewer relevant orbitals or with lower dimensionality) to refine this recipe and help discriminate between possible mechanisms. Much as the iron-pnictide superconductors have helped refine concepts proposed in the study of the cuprates such as cementing the importance of magnetism and demonstrating a tunability to the strength of electron-correlations in the normal state, a new family of non-centrosymmetric unconventional superconductors \AtTA\ (with \textit{A} = Na, K, Cs or Rb and \textit{TM} = Cr or Mo) with a structural motif of CrAs tubes have begun offering new insights into unconventional superconductivity, but importantly in a quasi-one-dimensional structure \cite{Basov2011,Allred2016,Bao2015a,Mu2018,Mu2018b,Tang2015r1,Zhao2018}. 

Though nascent, the study of these materials (specifically \KtCA ) has yielded helpful corroborations and refinements of our recipe. Neutron scattering, nuclear magnetic resonance, Raman scattering and density functional theory work have revealed both structural and magnetic instabilities - as seen in the pnictides \cite{Taddei2017t,Taddei2018,Zhi2015,Zhang2015,Wu2015,Jiang2015}. However, in \KtCA\ these orders break different crystal symmetries, show no coupling, and have temperature dependences suggesting that only the magnetic order couples to superconductivity \cite{Taddei2017t,Taddei2018}. Additionally, evidence of Luttinger liquid type physics, spin-triplet superconductivity and a nodal gap function have emerged indicating the broader novelty of this system \cite{Watson2017,Shao2018,Liu2016,Adroja2015,Sun2017,Yang2015}. Yet, the more comprehensive study of this family has been inhibited by its extreme air sensitivity and its seeming unsuitability for doping studies. While recently success was found in continuously varying the \textit{A}-site between K and Na, the effect is of internal pressure, ideally we would be able to charge dope and create phase diagrams exploring how the discovered instabilities behave as the Fermi-surface and potentially the superconducting transition are tuned \cite{Mu2019}.   

Promisingly, one member compound (\KoCA ) has been found which exhibits both a diminished electron count and stability in air.  \KoCA\ (or \lq 133\rq\ as opposed to \lq 233\rq\ for \KtCA ) is created by soaking the 233 stoichiometry in an ethanol bath \cite{Bao2015}. This process, according to previous reports, discretely removes one K from the structure increasing the crystal symmetry from \Psmt\ to \Pstm\ while crucially retaining the quasi-one dimensional CrAs tubes (albeit now with centrosymmetry) \cite{Bao2015}. Though the K content cannot be continuously controlled and so does not allow a full charge doped phase diagram, \KoCA\ could afford some insight as to what endmembers of such a phase diagram might look like. 

Yet, results from \KoCA\ (and the broader \AoCA\  with \textit{A} = K, Rb and Cs family) have proven surprisingly contentious. Initial reports suggested no superconductivity and spin-glass like magnetism \cite{Bao2015,Tang2015r1}. However, more recent work which performed a post-bath hydrothermal anneal in ethanol reported bulk superconductivity with non-magnetic behavior \cite{Mu2017,Liu2017}. Similarly, theory work has resulted in significant disagreement with reports of long range magnetic ground states, spin-density wave Fermi-surface nesting mediated superconductivity and more recently of a structural instability which precludes spin-fluctuation driven superconductivity \cite{Cao2015,Zhang2019,Xing2019}. 

An ostensible explanation for these conflicting results has been found in crystalline disorder - in Ref.~\onlinecite{Mu2017} it is argued that superconductivity arises in \KoCA\ upon post-treatment with a hydrogthermal anneal due to an improvement of the crystallinity. While in theory work, Ref~\onlinecite{Cao2015} suggests its predicted long range magnetic order is not seen experimentally due to disorder in the physical material and Ref~\onlinecite{Xing2019} suggests that similar to \KtCA\ the predicted structural instability may have no long range coherence \cite{Taddei2018}. In attempting to sort through these possibilities in the work reported here, we stumbled upon a different and quite intriguing origin to these discrepancies. 

\begin{figure}
	\includegraphics[width=\columnwidth]{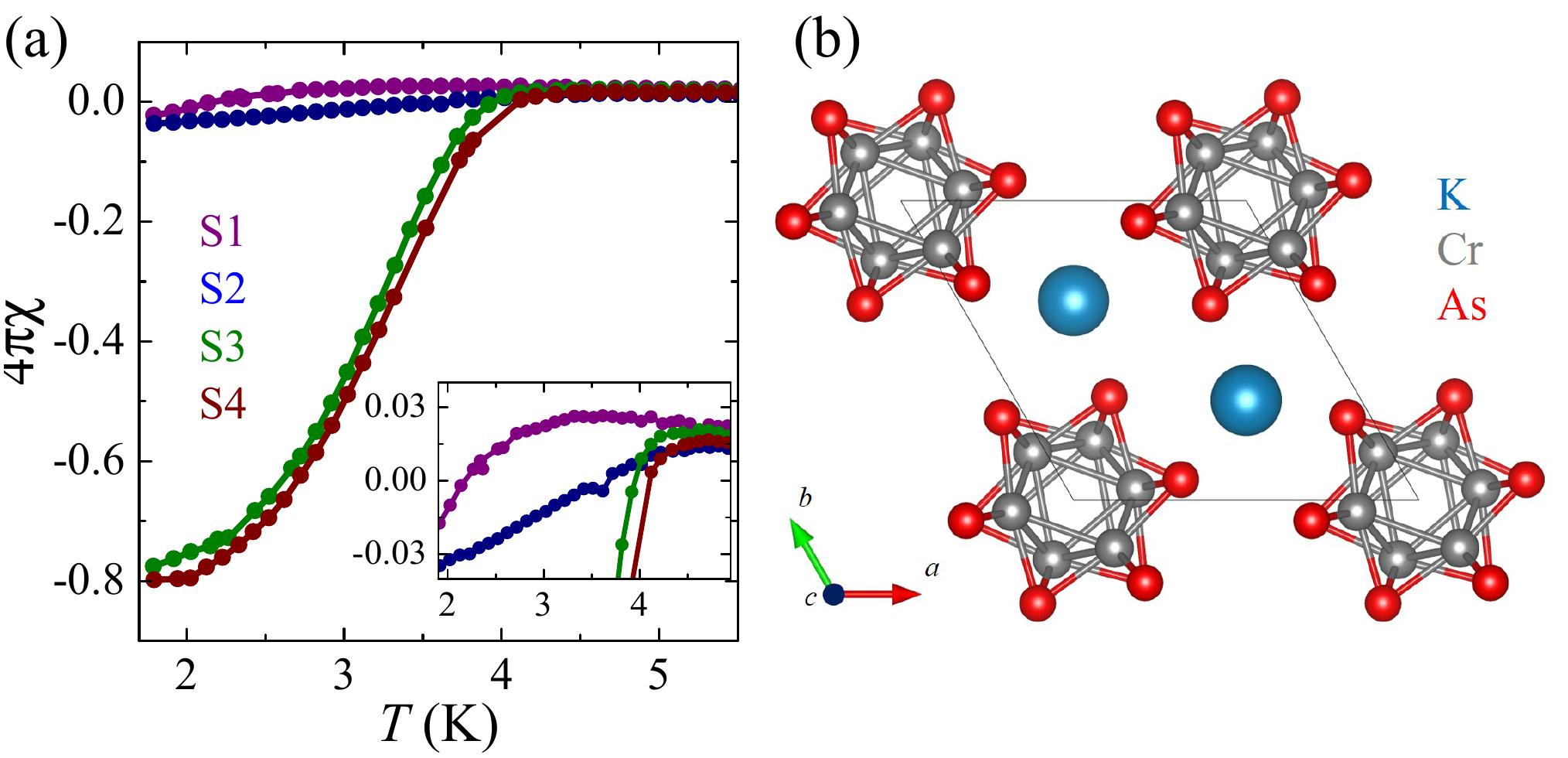}
	\caption{\label{fig:one} (a) Diamagnetic response measured plotted in unitless $4\pi \chi$ in a 10 Oe field under zero field cooled conditions. The inset is shows a close up of the onset of the transitions. (b) the previously reported crystal structure of \KoCA .  }	
\end{figure} 

Here we report a systematic study of \KoCA\ and its various reported growth techniques using neutron and x-ray diffraction together with Density Functional Theory (DFT). We reveal a previously missed effect of the ethanol bath deintercalation - H doping. Using structural solution methods we find that H intercalates into the CrAs tubes, partially charge compensating for the hole-doping effect of the deintercalated K. Furthermore, we show that the difference between non-superconducting spin-glass and non-magnetic superconducting samples is not in the sample crystallinity but is rather in the amount of H intercalated. Our DFT work shows that the intercalated \KHCA\ structure is stable, electron-doped relative to \KoCA\ and consequently has an electronic structure more similar to \KtCA\ than that calculated for \KoCA . Additionally, we find that the H stabilizes the \Pstm\ structure and places the compound proximate to an antiferromagnetic instability. These results lead to the exciting possibility of continuous tuning \KHCA\ from a spin-glass to a superconductor via charge doping with H - the long sought after tuning parameter. Furthermore this shifts interest to a new air stable composition opening up this system for widespread study. More contemplatively, it offers a realization of a frustrated magnetic to superconducting transition in a Q1D system which has long been suggested as a potential recipe for spin liquid physics \cite{Sigrist1994,Rice1993}.

\section{\label{sec:exp} Methods}

High quality powder samples of \KtCA\ were grown as described previously \cite{Bao2015,Taddei2017t}. To obtain samples of the 133 stoichiometry the deintercalation methods reported in \onlinecite{Bao2015} and refined in \onlinecite{Mu2017} were both used. Powder samples of \KtCA\ were placed in room temperature baths of high purity dehydrated ethanol for seven days before being rinsed thoroughly with ethanol and dried under vacuum. With the initial reaction complete, samples underwent one of three post treatments (i) left as grown, (ii) post-annealed under vacuum or (iii) annealed in a hydrothermal vessel with ethanol and then vacuum annealed. The powders obtained from all reactions were of gray color and were air stable on the order of days. The crystal structure using the initially reported stoichiometry is shown in Fig.~\ref{fig:one}(b). Further details on the synthesis procedures are included in the supplemental materials (SM)\cite{SM}.   

\begin{figure*}
	\includegraphics[width=\textwidth]{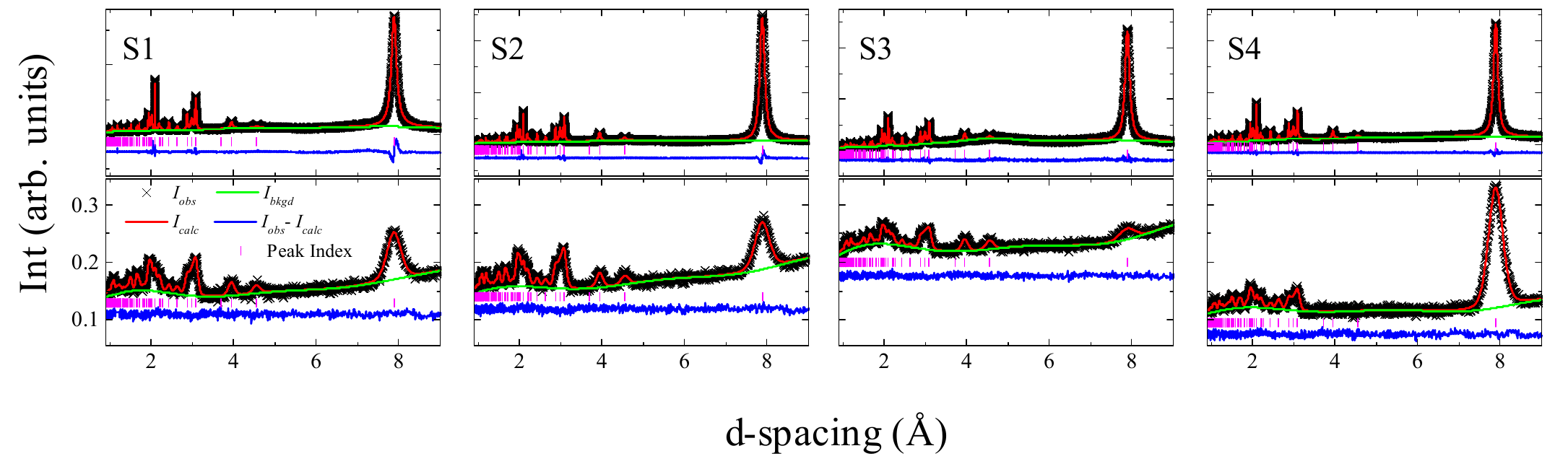}
	\caption{\label{fig:two} Room temperature x-ray (top row) and neutron (bottom row) powder diffraction patterns for the four \KoCA\ samples plotted with the results of Rietveld refinements for H and D containing models. The neutron diffraction data was collected using identical sample mass, neutron exposures and sample containers and are thus the scattered neutron intensity is directly comparable between samples.}	
\end{figure*}

Neutron powder diffraction (NPD) data were collected using the HB-2A powder diffractometer and the NOMAD total scattering instruments of Oak Ridge National Laboratory's High Flux Isotope Reactor and Spallation Neutron Source \cite{Calder2018}. For the latter the data were reduced using the instrument developed software \cite{McDonnell2017, Neuefeind2012}. High resolution synchrotron x-ray powder diffraction (XRD) data were collected at beamline 11BM-B of the Advanced Photon Source at Argonne National Laboratory.  Structural analysis was performed using the Rietveld method as implemented in the FullProf, GSAS and EXPGUI software suites (see SM for more details) \cite{Rodriguez-Carvajal1993,Larson2004,VonDreele1982,Finger1994, SM}.  

Density Functional Theory (DFT) calculations were performed using projector-augmented plane-wave method, as implemented in the VASP code, with the Perdew-Burke-Ernzerhof generalized gradient approximation and checked using the general potential linearized augmented planewave method as implemented in the WIEN2k code \cite{Kresse1999, Kresse1996,Perdew1996, Singh2006, Blaha2001}. For the basis set, an energy cutoff of 400 eV was used and a Brillouin zone was composed of $4\times 4 \times 10$ \textit{k}-point sampling. Starting from the experimentally determined lattice cell, the internal atomic positions were relaxed subject to hexagonal symmetry with the lattice parameters held at the experimental values. The phonons were calculated using the PHONOPY code \cite{Togo2015}. For more details on the calculations see the SM \cite{SM}.


\section{\label{sec:disc} Results and Discussion}

To systematically compare the thermodynamic and crystallographic properties of samples grown using different techniques, we synthesized samples following the methodologies reported by Bao \textit{et al.,} in Ref.~\onlinecite{Bao2015} and Mu \textit{et al.,} in Ref.~\onlinecite{Mu2017}. Thus three samples were used for initial characterization: (S1) as grown from an ethanol bath, (S2) with a post-bath vacuum anneal and (S3) with both post-bath hydrothermal and vacuum anneals (a fourth sample (S4) was also grown and will be discussed later in this text). Susceptibility measurements (Fig.~\ref{fig:one} (a)) show  we were successful in reproducing the previously reported change in properties with S1 and S2 being only trace-superconducting with signatures of spin-glass behavior (identified here as in Ref.~\onlinecite{Bao2015} by the low temperature divergence of the inverse susceptibility from linearity and a split between field-cooled and zero-field cooled measurements, see SM) and S3 showing bulk superconductivity.

Previously, the change from spin-glass to superconductor was attributed to the post-bath annealings' effects on the crystallinity of the sample. Ostensibly, the more crystalline annealed samples hosted superconductivity which was otherwise destroyed by disorder in the as-grown samples \cite{Mu2017}. To test this assertion NPD and XRD data were collected and are shown along with the best fit results of Rietveld refinements in Fig.~\ref{fig:two}. We note that even visually, the main qualitative difference between the samples is not in the peak shape but rather is in the background to signal ratio. Specifically, in the NPD data the bulk superconducting S3 has a remarkably larger background to signal ratio than any other sample. 

\begin{figure}
	\includegraphics[width=0.65\columnwidth]{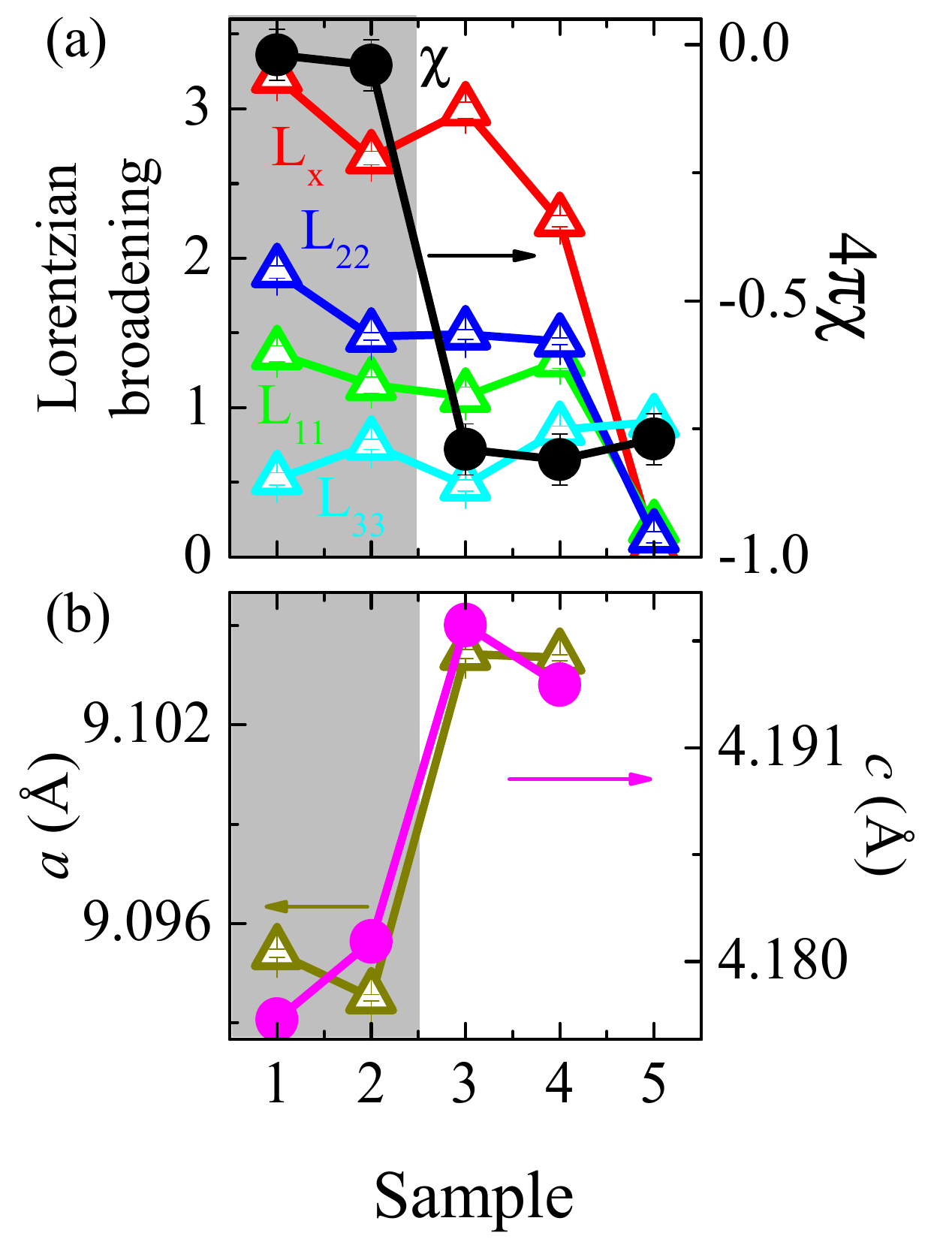}
    \caption{\label{fig:three} (a) Lorentzian broadening terms in the GSAS profile fitting function including three terms for micro-strain along the unit cell directions (L$_{11}$, L$_{22}$ and L$_{33}$) and one term (L$_{x}$) for particle size broadening. This latter is proportional to the inverse of the particle size. The superconducting volume fraction determined at 1.5 K is over-plotted with sample 5 denoting the pure 233 compound. (b) the refined \textit{a} and \textit{c} lattice parameters. Throughout the figure data with scale on the left axis are plotted with open triangles while data with scale on the right axis are plotted as filled circles. The grey shaded region denotes samples with only trace superconductivity.}	
\end{figure}

To qualitatively check the crystallographic properties the NPD and XRD were used together for Rietveld refinements. From this analysis, the lattice parameters and measures of strain, crystallinity and crystal size can be assessed (plotted as a function of sample number along with the 233 compound labeled as sample 5 in Fig.~\ref{fig:three}). Considering the extracted anisotropic microscopic strain parameters ($L_{nn}$ where $n = 1,2,3$ correspond to the $a, b \text{ and } c$ lattice directions respectively) and  $L_{x} $ (which is proportional to the inverse of the crystallite size), little difference is seen in these peak broadening terms between samples indicating similar crystalline properties. Compared to the broadening parameters determined in previous work on the 233 compound (Ref.~\onlinecite{Taddei2017t}), it is seen that generally the deintercalation process leaves the 133 material with poor crystallinity, as reported. However, it is not significantly improved upon either vacuum or hydrothermal annealing, in contrast with previous reports which did not compare qualitative measures of the crystallinity \cite{Mu2017}. 

Looking at the crystal parameters rather than the crystallinity a clear trend relative to the superconducting volume fraction is seen (Fig.~\ref{fig:three} (b)). Both the $a$ and $c$ lattice parameters dilate between the trace (S2) and bulk (S3) superconducting samples by 0.1 and 0.4 \%\ respectively leading to a $>0.4\%\ $ change in the unit cell volume (Fig.~\ref{fig:three} (b)). On the other hand, between S1 and S2 no such change is seen with only a small $\sim$ 0.1\%\ change in volume and no significant change in the superconducting volume fraction. Both of these behaviors are unexpected: if the post-annealing treatments improve the crystallinity the profile peak broadening terms should decrease and the unit cell relax - our data contradict both of these expectations \cite{Forrest2016,Becht2008}. Therefore, we aimed to identify a different cause to the change in properties between the as-grown/vacuum-annealed samples and the hydrothermally annealed samples.

Noting that the background to signal increase is only seen in the NPD and is exacerbated by the hydrothermal anneal, we posit the presence of H in the samples. H is nearly invisible to x-rays but has a large incoherent cross-section to neutrons ($\sigma_{H_{Inc}} \sim 80.26$ $b$) \cite{Sears2006}. This manifests as a large background which increases at lower scattering angle (or high d-spacing) and would explain the difference between the XRD and NPD patterns \cite{Temleitner2015}. As such, any H in \KoCA\ would be missed in XRD and energy dispersive x-ray spectroscopy indicating why such a feature has not been seen in previous structural work \cite{Bao2015, Mu2017, Liu2017, Tang2015r1}. One could propose several mechanisms which would introduce H to the samples during the two ethanol baths: as a minority organic compound not removed with ethanol washing, substituted onto the K site or intercalated into the 133 structure. Of these scenarios the latter two provide an explanation for the increase in background upon the hydrothermal anneal assuming the process either forces more H into the material through higher temperature, pressure or a longer exposure to ethanol. 

To discriminate between these scenarios, a sample of 133 was grown following the procedure of S3 but using fully deuterated ethanol in both the room temperature bath and hydrothermal anneal. The resulting sample (S4) shows simlar superconducting properties to S3 (Fig.~\ref{fig:one}) indicating that the use of deuterated ethanol does not significantly change the sample properties or reaction pathway. Deuterium (D) does not have the large incoherent cross section of H ($\sigma_{D_{Inc}} \sim\ 2.05$ $b$) and so does not add as much background scattering. Furthermore, it has a large coherent cross-section of $\sigma_{D_{Coh}} \sim\ 7.6$ \textit{b} making it a strong scatterer suitable for structural model refinements (the coherent cross-sections of K, Cr, As and H are 1.69, 1.66, 5.44 and 1.76 \textit{b} respectively) \cite{Sears2006}. Therefore, if H does intercalate into the structure or substitute onto the K site using D would allow for structure solution and composition determination through changes in peak intensities, else if the H background comes from remnant organic compounds the background will be reduced in S4 but no other changes should occur.

The XRD and NPD of S4 are shown in Fig.~\ref{fig:two}. While the XRD looks similar to those of the other samples, the NPD exhibits a nearly 50\%\ reduction in background (compared to S3) as well as a change in peak intensities where peaks at 3.9 and 4.6 $\mathrm{\AA}$ (the (020) and (110) reflections respectively) become background equivalent. The background reduction confirms the presence of H in the samples (rather than for  instance the background arising from amorphous parts of the sample) while the change in peak intensities between the samples grown with H and D indicates that the H(D) is incorporated into the 133's crystal structure rather than existing as remnant organic compounds from the ethanol bath. 

To determine where D is incorporated, Rietveld analysis was performed using the NPD. Since the D containing samples show a reduction in peak intensity but not the appearance of new peaks the original \Pstm\ crystal symmetry was used and Rietveld refinements were performed trying models containing D on all of \Pstm 's special Wyckoff positions. Of these, only models with D on the $2b$ (0,0,0) site reproduced the observed intensities as shown in Fig.~\ref{fig:two} leading to high quality fits with satisfactorily small residuals (Table~\ref{tab:one}). We note that the new reflection conditions of these sites do not forbid either the (020) or (110) reflections by symmetry instead they become accidentally forbidden.   
 
\begin{table}
	\caption{\label{tab:one}Crystallographic properties of samples obtained from Rietveld refinements using both the XRD and NPD data at 300 K.}
	\begin{ruledtabular}
		\begin{tabular}{lcllll}
		\multicolumn{2}{c}{} & \multicolumn{1}{c}{S1} & \multicolumn{1}{c}{S2} & \multicolumn{1}{c}{S3} & \multicolumn{1}{c}{S4}\\
		\hline
		Space group   &       & \multicolumn{4}{c}{$P6_3/m$}   \\
		$R_{wp}$      &       & 4.13\%    & 3.56\%     & 2.46\%    & 3.36\%    \\
		$a$ ($\AA$)   &       & 9.0948(1) & 9.0939(1)  & 9.1042(1) & 9.1041(1) \\
		$c$ ($\AA$)   &       & 4.1770(3) & 4.18090(3) & 4.19733(1)& 4.19426(2)\\
		$V$ ($\AA^3$) &       & 299.22(1) & 299.43(1)  & 301.29(1) & 301.06(1) \\
		              &       &           &            &           &           \\
		K($2c$)       &       &           &            &           &           \\
                      & $U$   & 0.020(6)  & 0.010(5)   & 0.017(7)  & 0.018(1)  \\   		
		Cr($6h$)      &       &           &            &           &           \\
		              & $x$   & 0.1488(2) & 0.1498(1)  & 0.1520(2) & 0.1512(1) \\
		              & $y$   & 0.1850(2) & 0.1867(1)  & 0.1887(2) & 0.1885(1) \\
		              & $U$   & 0.009(1) & 0.004(1)    & 0.003(1)  & 0.005(2)  \\ 
		As($6h$)      &       &           &            &           &           \\
		              & $x$   & 0.3438(1) & 0.3430(1)  & 0.3459(1) & 0.3461(1) \\
		              & $y$   & 0.0672(1) & 0.0672(1)  & 0.0688(1) & 0.0697(1) \\
		              & $U$   & 0.014(2)  & 0.013(2)   & 0.011(1)  & 0.009(2)  \\ 
		H($2b$)       &       &           &            &           &           \\
			          & $occ.$& 0.35(1)   & 0.34(1)    & 0.71(3)   & 0.65(1)   \\ 
                      & $U$   & 0.015     & 0.015      & 0.015     & 0.015(3)  \\ 
		              
		\end{tabular}
	\end{ruledtabular}
\end{table}
 
\begin{figure}
	\includegraphics[width=\columnwidth]{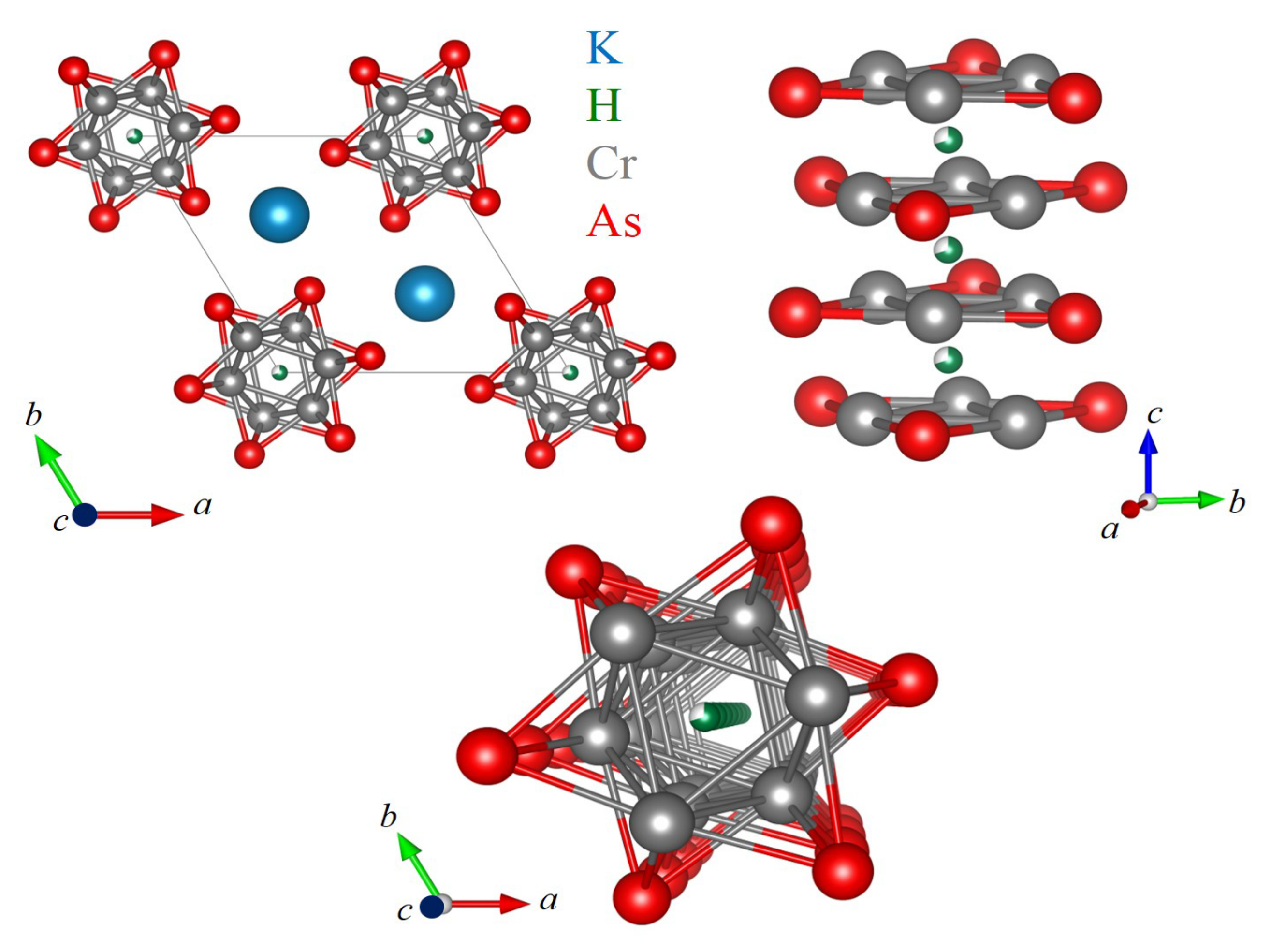}
	\caption{\label{fig:four} \KHCA\ structure as viewed along the \textit{c}-axis, perpendicular to the \textit{c}-axis and along the CrAs tubes. K, H, Cr and As atoms are denoted by turquoise, dark green, gray and red spheres respectively. The partial occupancy of the H site is indicated by a wedge of white in the H atom surface.}	
\end{figure}

The new structural model is shown in Fig.~\ref{fig:four}. In it, D is intercalated into the center of the CrAs tubes with the $2b$ site located between the stacked CrAs layers creating a chain of D along the center of each tube. We find no evidence of K vacancies or of D substitution on the K site - changes neutron scattering is very sensitive to due to the considerable difference in the neutron scattering cross sections of K and D. However, our refinements show the D site is only partially occupied with a refined occupancy of 65\% .  

With the location of the intercalated D known, we can attempt Rietveld refinements using similar models for S1, S2 and S3, this is possible since despite its large incoherent cross-section H also has a coherent cross-section which is on the order of both K and Cr. Starting with S3, our refinements show remarkable agreement with S4, with a similar occupancy of $\sim 71$\% . This corroborates our assumptions from the susceptibility that the synthesis procedure and resulting material is insensitive to the change from H to D. On the other hand, S1 and S2 are found to have significantly less H than S3 and S5. In both S1 and S2 the $2b$ site refines to have $\sim 35\%$ occupancy. As all samples show intercalated H - even the non-bulk superconductors - we revise the chemical formula of \KoCA\ to \KHCA\ to explicitly reflect the H content and its variability.    

These results elucidate the effects of the different reaction stages and show each step to be quite reproducible. During the room temperature ethanol bath, K is deintercalated from the 233 starting material as reported previously however, as our analysis shows some H from the bath also intercalates into the CrAs tubes at positions between the CrAs layers. Whether this occurs to charge compensate for the deintercalated K is not known and should be the interest of future experimental and theoretical work - although this may be indicated by the inability to synthesize \KoCA\ directly from sintering \cite{Bao2015, Mu2017}.  The similarities both in crystallinity and H content of S1 and S2 suggest that vacuum annealing does not have a significant effect at this stage, it does not improve the sample quality and causes no venting of the intercalated H. This latter result suggests the stability of \KHCA\ at least up to the low annealing temperatures (373 K) used in this work. Upon annealing in a hydrothermal ethanol bath, more H intercalates into the structure bringing the concentration from $\sim 35\% $ to $\sim 65 - 71\% $.    

It is these last two effects which appear to induce superconductivity in \KHCA . As was discussed earlier no change in crystallinity is seen between bulk and trace superconducting samples but rather a previously unexplained change in the lattice parameters. We we can now ascribe this to an increase in the amount of H intercalated into the CrAs tubes. Beyond causing the unit cell to expand, the intercalated H should also affect the charge count of the CrAs tubes. However, the chemistry of H is incredibly flexible and it is not evident from the diffraction data how the H contributes electronically therefore, we turn to DFT to help uncover the role of H in the structure \cite{Gupta2005, Khan1976}. 

\begin{figure}
	\includegraphics[width=\columnwidth]{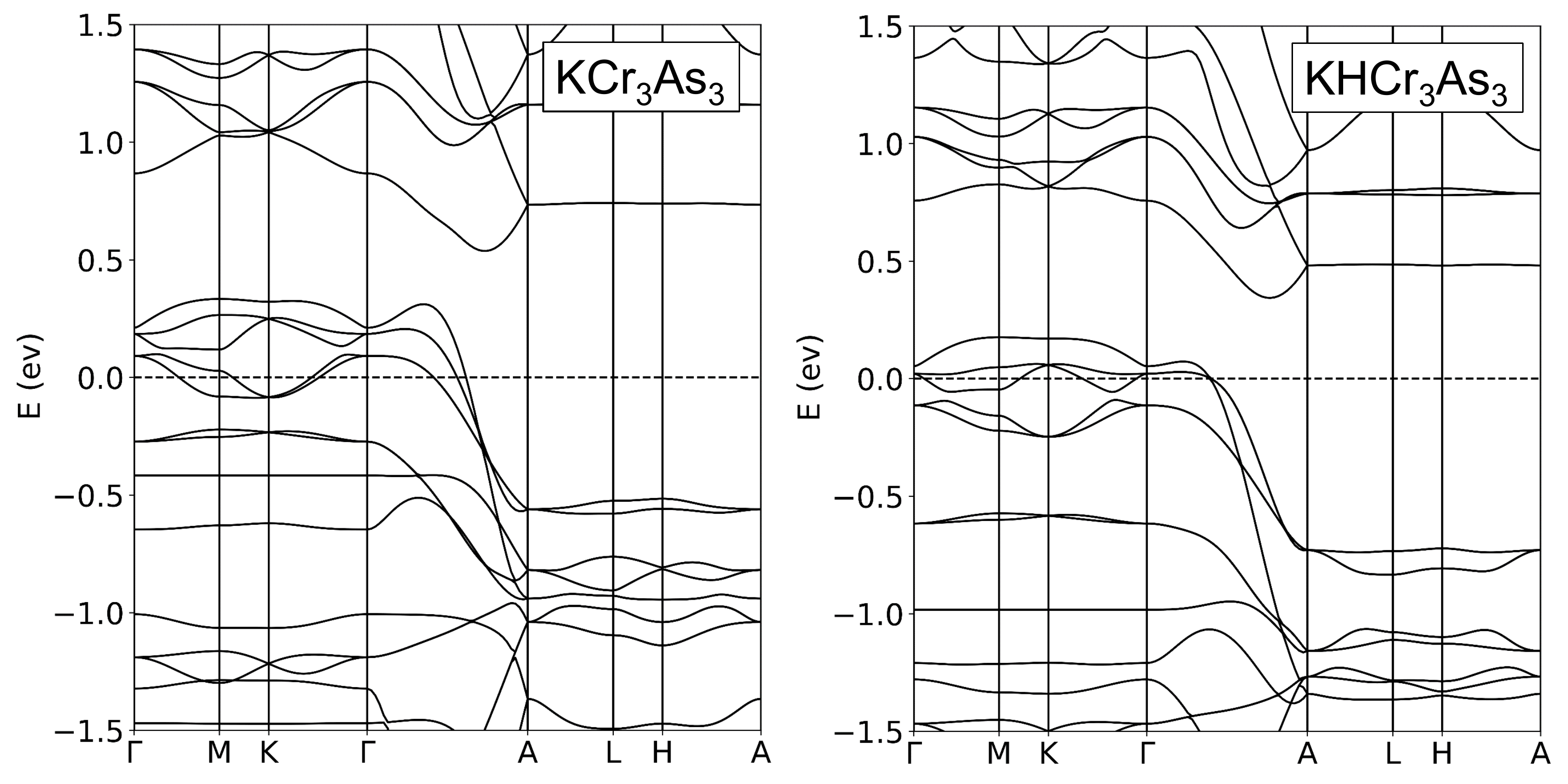}
    \caption{\label{fig:five}  DFT calculated band structure for (left) the \KoCA\ and (right) \KHoCA\ stoichiometries.}	
\end{figure}

\begin{figure}
	\includegraphics[width=0.3\columnwidth]{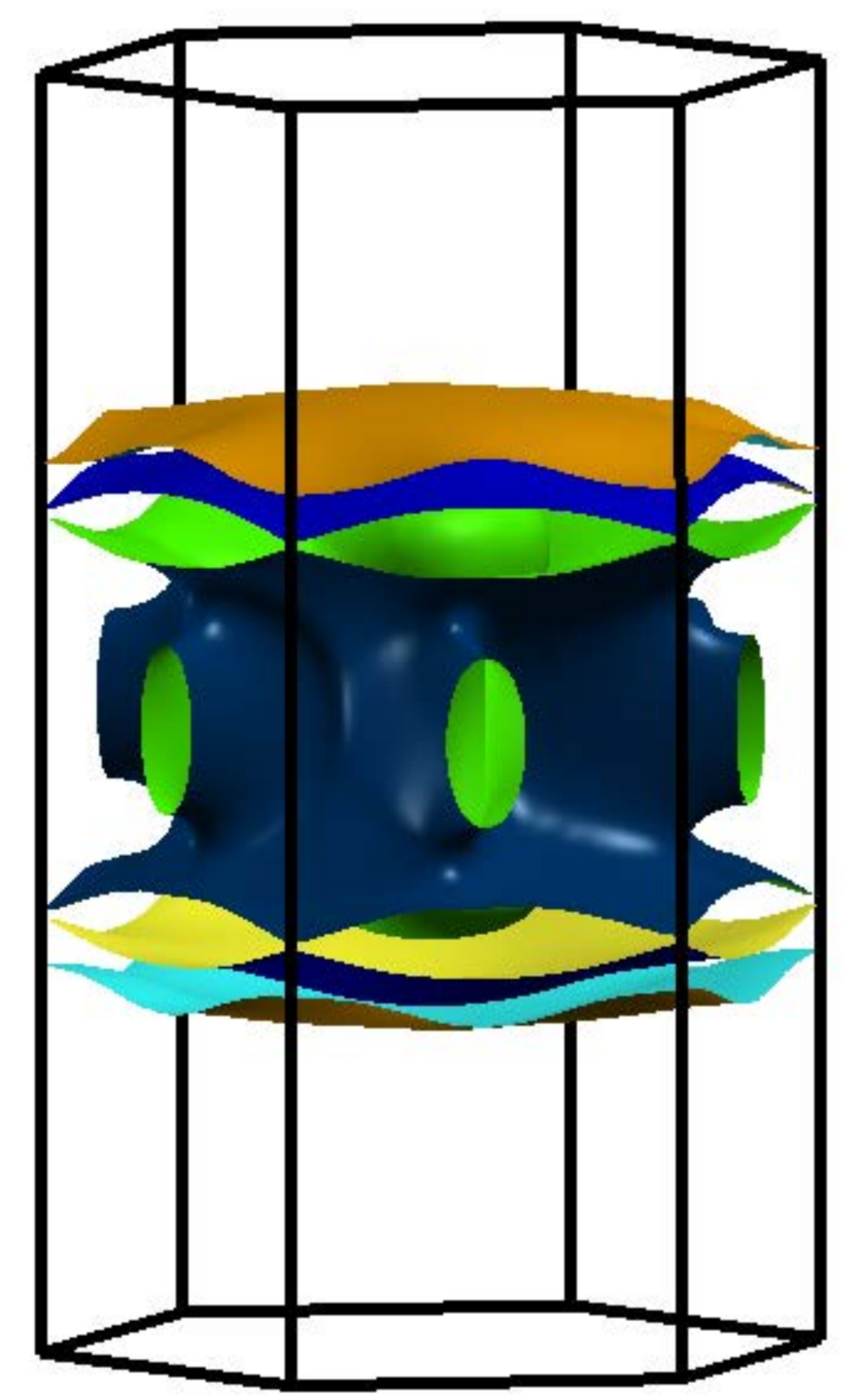}
    \caption{\label{fig:six}  DFT calculated Fermi surface of \KHoCA\.}	
\end{figure}

To study \KHCA 's energetics and electronic structure, DFT calculations were performed using both the theoretical $x = 0$ and $x =1$ stoichiometries. Starting with the energetics, calculations for \KHoCA 's formation energy relative to \KoCA\ and H$_2$ gas were performed. These found a negative formation energy of 92 kJ/mol H$_{2}$ - similar to that of other stable hydrides such as MgH$_{2}$ - indicating the intercalated structure is energetically favorable (see SM for more details) \cite{SM}. Next, we investigated the electronic contribution of H by calculating \KHoCA 's band structure (Fig.~\ref{fig:five}). We found that the intercalated H raises the Fermi energy ($E_{F}$)and causes only modest distortions to the bands near $E_{F}$. Therefore, we can say that in \KHoCA\ the H has metallic bonding and acts as an electron donor. As shown, it increases the electron count such that the electronic structure (and Fermi surface Fig.~\ref{fig:six}) are more similar to \KtCA\ than the pure \KoCA\ ostensibly indicating similar electronic behavior. 

Inspired by previous reports of structural and magnetic instabilities we also performed calculatinos studying the phonon spectrum and proximity to magnetic instabilities \cite{Xing2019, Cao2015,Feng2019,Zhang2019}. considering the predicted structural instability, our calculations show neither the unstable phonon modes seen in \KoCA\ or \KtCA \cite{Taddei2018, Xing2019}. Furthermore, we find relatively high energy nearly dispersionless H modes and thus suggest that H removes the unstable phonon modes leading \KHoCA\ to be structurally stable \cite{Rowe1974}. Finally, we consider the hydride's proximity to a magnetic instabilities. Previously, \KoCA\ has been suggested to either have a magnetic ground state which is frustrated by disorder, or no magnetic ground state due to a structural instability which removes favorable nesting conditions \cite{Cao2015,Xing2019}. For \KHoCA\ we find weak itinerant instabilities to both ferromagnetic and antiferromagnetic A-type magnetic orders with a slight preference for the latter. However, we also find both to have itinerancy and low energy scales suggesting that the magnetic order could easily be suppressed to a paramagnetic state with spin-fluctuations by quantum fluctuations such as the case of Ni$_3$Ga \cite{Aguayo2004}. This last result is quite interesting as it restores the possibility of a spin-driven pairing mechanism which had previously been discounted in Ref.~\onlinecite{Xing2019}.     

These DFT and experiment results suggest rather provocatively that superconductivity arises in \KHCA\ upon electron doping, a process which until now has been achieved unintentionally during the hydrothermal anneal. In this scenario, the underdoped \KHtfCA\ is a non-superconducting spin-glass phase as reported in Bao. \textit{et al.}'s initial report \cite{Bao2015}. Upon electron doping to \KHsnCA\ via hydrothermal annealing, the spin-glass phase is suppressed and the material becomes a bulk superconductor with \Tc\ $\sim 5$ K as reported by Mu \textit{et al.,} \cite{Mu2017}. This has multiple significant implications: (i) the intercalated CrAs systems inhabit a familiar phase space where superconductivity arises out of a suppressed magnetic state, (ii) the possibility of creating the previously elusive charge doped phase diagram for the intercalated CrAs for not just \KHCA\ but also the Rb and Cs analogues all of which are grown in ethanol and show the spin-glass to superconductor crossover, (iii) the need to revisit previous \textit{ab initio} calculations on the \AoCA\ materials which were based on the hydrogen free 133 stoichiometry, (iv) an avenue to study how superconductivity arises from a frustrated magnetic state in a quasi-one-dimensional system by charge doping a concept which has long held interest for spin- and Luttinger liquid type physics \cite{Sigrist1994,Rice1993}. As an additional point of interest, one of the significant impediments to the study of the intercalated CrAs materials to this point has been the initial 233 stoichiometry's extreme air sensitivity, with this work, we show that the 133 version offers a clear path forward to study interactions between magnetic and superconducting orders while also being air stable alleviating this barrier. Our results layout a rich path of future work, not least of which will be to gain fine control of the H content through the synthesis procedures - work which these authors have already begun to early positive indications. 

\section{\label{sec:con} Conclusions}

In conclusion, we have shown via neutron and x-ray diffraction that during the wet chemistry K deintercalation of \KtCA , H intercalates into the structure forming a partially occupied H chain centered in the CrAs tubes. We find that it is the concentration of H and not the crystalline quality of the material which determines the superconducting properties with the reported synthesis methods producing the non-bulk-superconducting spin-glass \KHtfCA\ and the bulk-superconducting metallic \KHsnCA . Density functional theory analysis shows that the \KHoCA\ structure is energetically very favorable and that the intercalated H acts as a electron donor. The resulting electronic structure is similar to that of the \KtCA\ compound, does not have the structural instability predicted for \KoCA\ and has weak magnetic instabilities. These results suggest the need to revisit much of the theoretical work on these materials as well as multiple new research avenues to understand unconventional superconductivity and magnetic interactions in these novel quasi-one-dimensional systems.

\begin{acknowledgments}
The part of the research that was conducted at ORNL’s High Flux Isotope Reactor and Spallation Neutron Source was sponsored by the Scientific User Facilities Division, Office of Basic Energy Sciences, US Department of Energy. The research is partly supported by the U.S. Department of Energy (DOE), Office of Science, Basic Energy Sciences (BES), Materials Science and Engineering Division. Work at the University of Missouri is supported by the U.S. DOE, BES, Award No. DE-SC0019114. The authors thank J. Neuefeind and J. Liu for help with data collection on NOMAD.
\end{acknowledgments}


%

\end{document}


\preprint{APS/123-QED}

\title{Supplemental Material for: \\ Tuning from Frustrated Magnetism to Superconductivity in Quasi-One-Dimensional \KoCA\ Through Hydrogen Doping}

\author{Keith M. Taddei}
\email[corresponding author ]{taddeikm@ornl.gov}
\affiliation{Neutron Scattering Division, Oak Ridge National Laboratory, Oak Ridge, TN 37831}
\author{Liurukara D. Sanjeewa}
\affiliation{Materials Science and Technology Division, Oak Ridge National Laboratory, Oak Ridge, TN 37831} 
\author{Bing-Hua Lei}
\affiliation{Department of Physics and Astronomy, University of Missouri, Missouri 65211, USA}
\author{Yuhao Fu}
\affiliation{Department of Physics and Astronomy, University of Missouri, Missouri 65211, USA}
\author{Qiang Zheng}
\affiliation{Materials Science and Technology Division, Oak Ridge National Laboratory, Oak Ridge, TN 37831}
\author{David J. Singh}
\affiliation{Department of Physics and Astronomy, University of Missouri, Missouri 65211, USA}
\author{Athena S. Sefat}
\affiliation{Materials Science and Technology Division, Oak Ridge National Laboratory, Oak Ridge, TN 37831}
\author{Clarina dela Cruz}
\affiliation{Neutron Scattering Division, Oak Ridge National Laboratory, Oak Ridge, TN 37831}

\date{\today}

\maketitle

\section{\label{sec:exp} Sample Synthesis}

To obtain \KoCA , first \KtCA\ was synthesized following the methodology described in Ref.~\onlinecite{Taddei2017t} the resulting 2 g pellet was ground and characterized by x-ray diffraction and susceptibility measurements to confirm it as single phase and bulk superconducting. The sample was then separated into four 0.5 g samples (labeled S1, S2, S3 and S4) for use in the different deintercalation reactions. S1 and S3 were used to replicate the procedures reported in Refs.~\onlinecite{Bao2015} and \onlinecite{Mu2017} respectively. S2 was used to check superconductivity as a function of crystallinity by performing a vacuum post-anneal on a sample following the single bath procedure of Ref.~\onlinecite{Bao2015} and S4 was used to check for incorporation of H into the structure by following the procedure of Ref.~\onlinecite{Mu2017} but with high purity deuterated ethanol (E-D6) used in both the room temperature and hydrothermal baths. 

In the first step all samples followed the same procedure (save for using E-D6 with S4). Each 0.5 g sample was placed in a 5 mL glass vial with 1 mL of pure ethanol (or E-D6) in a high purity He atmosphere glovebox. To slow the loss of ethanol, parafilm was wrapped over the top of each vial and lightly perforated to allow the H$_2$ gas to escape. When the ethanol was added an immediate reaction was visible with a rapid bubbling in the \KtCA\ powder. This rapid bubbling ceased within the first several hours. Each sample was left to sit in the ethanol bath for 7 days. The samples were regularly checked and as ethanol evaporated more was added to maintain ~1 mL of ethanol in the vials at all times. Each time more ethanol was added the samples were agitated to ensure all sample was exposed to the bath. In total 2 mL of ethanol was used for each sample. After 7 days the samples were rinsed twice each with 1 mL of ethanol (E-D6) used in each rinse. The samples were then placed under vacuum for 30 mins to ensure no remaining ethanol. The obtained powders were dark gray and easily agitated. 

\begin{table}
	\caption{\label{tab:one}Synthesis soak and anneal times}
	\begin{ruledtabular}
		\begin{tabular}{ccccc}
		\multirow{2}{*}{Sample } & \multicolumn{1}{c}{Ethanol} & \multicolumn{1}{c}{Hydrothermal} & \multicolumn{1}{c}{Vacuum} & \multirow{2}{*}{Deuterated}\\
		 & \multicolumn{1}{c}{bath (hrs)} & \multicolumn{1}{c}{bath (hrs)} & \multicolumn{1}{c}{anneal (hrs)} & \\
		\hline
		S1  & 168 &   0 &  0 & N \\
		S2  & 168 &   0 & 12 & N \\
		S3  & 168 & 100 & 12 & N \\
		S4  & 168 & 100 & 12 & Y \\
		\end{tabular}
	\end{ruledtabular}
\end{table}

At this stage the different samples underwent different post-bath treatments. S1 saw no further treatment. S2 was post-annealed in an alumina crucible in an evacuated quartz tube for 12 hrs at 373 K. S3 and S4 were loaded into autoclaves with 2 mL of ethanol and E-D6 respectively. The autoclaves were loaded in the He atmosphere glovebox. S3 and S5 were then sintered at 353 K for 100 hrs after which they were rinsed and left under vacuum for 30 mins.        

\begin{figure}[h]
	\includegraphics[scale=0.5]{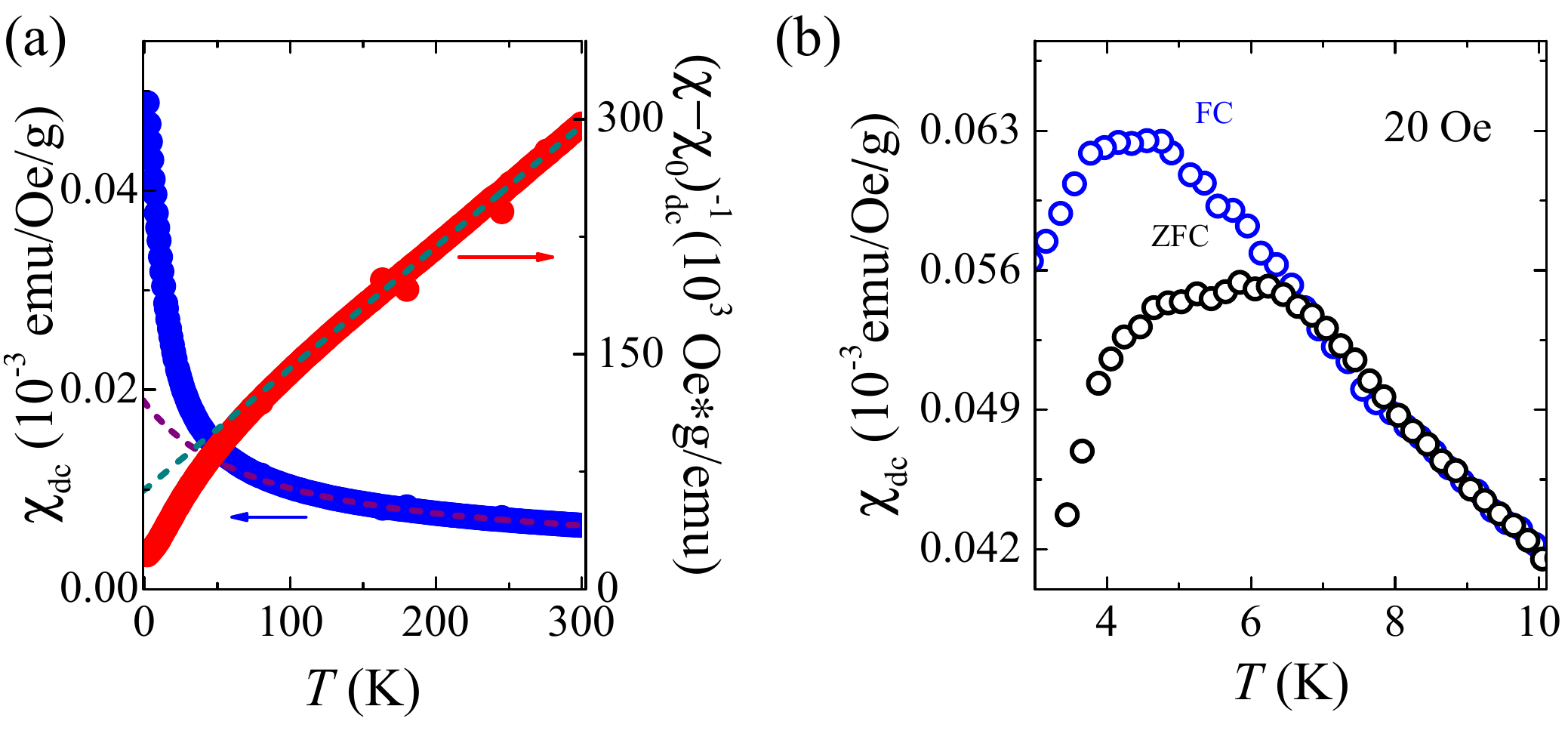}	
	\caption{\label{fig:sone} (a) Magnetic susceptibility and inverse susceptibility for S1 plotted with modified Curie-Weiss fits to show the divergence from Curie-Weiss behavior and ostensible spin-glass transition. (b) Low temperature susceptibility collected using field-cooled and zero-field-cooled procedures, showing an enhanced ferromagnetic signal under field-cooling similar to that reported in Ref.~\onlinecite{Bao2015}. }	
\end{figure}   

The four obtained samples were then checked using x-ray diffraction and magnetic susceptibility measurements. All samples were found to be single phase with similar crystalline properties (as discussed in the main text.) Samples S1 and S2 showed only trace superconductivity with magnetic susceptibilities similar to those reported in Ref.~\onlinecite{Bao2015} and identified as indicative of spin-glass physics (Fig.~\ref{fig:sone}.)

\newpage

\section{\label{sec:GSAS} Neutron and X-ray Diffraction and Rietveld Refinements}


Powder diffraction patterns were collected on the HB-2A and NOMAD diffractometers of the High Flux Isotope Reactor and Spallation Neutron Source respectively of Oak Ridge National Laboratory. For the former the 1.54 \AA\ monochromator setting was used with no pre-sample collimation to maximize flux on sample. Patterns were collected between 300 and 2 K with 4 hrs count time. Only a preliminary sample grown following the same procedure as S1 was measured at HB-2A. It was loaded in an aluminum sample can under He atmosphere. Using the NOMAD diffractometer, data was collected at 300 and 100 K with 1 and 2 hr collection times respectively. All samples were loaded in glass capillaries for use with the instrument sample \lq shifter\rq\ sample changer. 

X-ray powder diffraction was collected using 11BM-B the high resolution powder diffractometer of Argonne National Laboratory's Advanced Photon Source. Samples were loaded into annular kapton capillaries and sealed on the ends with superglue. Measurements were taken at 300 and 100 K using 11BM-B's nitrogen cryostream with collection times of 1 hr per scan and an incident beam of 0.4145 \AA .     

Rietveld refinements were performed using the GSAS and EXPGUI software suites \cite{Toby2001,Larsen2015}. The time-of-flight profile function TYPE-3 of GSAS was used to model the peak profile via convoluted back-to-back exponentials convoluted with a psuedo-Voigt \cite{VonDreele1982}. Data from 11BM and NOMAD were used to perform joint refinements with banks 1 and 2 of NOMAD (using d-spacing coverages of $\sim 0.5 - 10$\AA\ and $\sim 0.5 - 5$ \AA\ respectively) used together with the 11BM data. By combining the two low scattering angle banks from NOMAD we were able to both monitor low scattering angle peaks (which arise from the 100 reflection) using bank 1 and have moderate resolution with bank 2.  All data was modeled well by a single phase using the reported \Pstm\ space group. We report the extracted crystallographic properties in the main text. 

\begin{figure}[h]
	\includegraphics[scale=0.5]{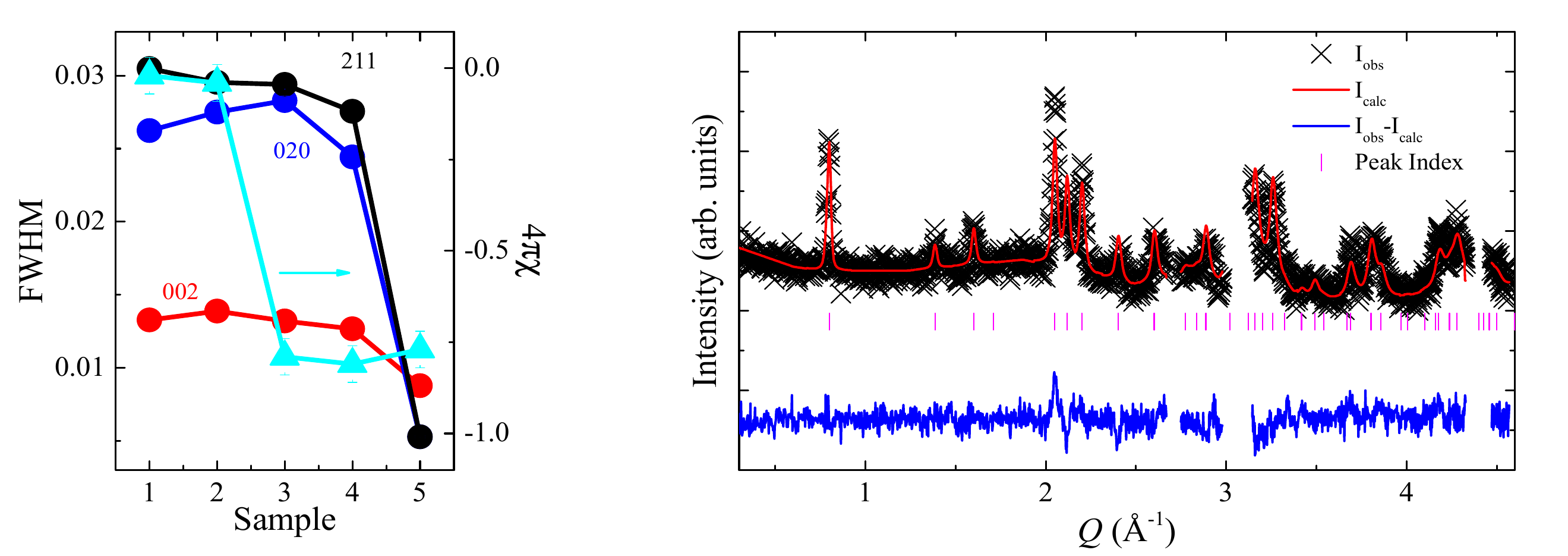}	
	\caption{\label{fig:stwo} (left) Full width at half max (FWHM) for a selection of diffraction peaks for samples S1, S1, S3, S4 and the 233 compound. Peaks from the high resolution 11BM-B data were used in the fitting. Results are plotted along with the superconducting volume fraction. (right) Neutron powder diffraction pattern collected on a preliminary sample grown similar to S1 using the HB-2A diffractometer. Fitting results in nearly identical parameters to those reported in the main text for S1. Gaps in pattern are regions excluded due to Al peaks coming from the sample can.}	
\end{figure} 

To confirm our analysis of the refined crystalline strain and size parameters we also looked directly for differences in the peak widths (Full width at half max FWHM) in the raw diffraction data (Fig.~\ref{fig:stwo}). Looking at the 002, 020 and 211 reflections (for their relative isolation in \textit{Q}) we see little change between the 4 133 compounds and no significant change between the superconducting and non-superconducting samples. Interestingly, we see that the largest difference in crysatllinity between the 133 and 233 samples is in-plane with the 020 peak being significantly broader in the 133 samples while the 002 peak only shows a modestly increased FWHM.

As a check of our refinement results for the H position and occupancy, in Fig.~\ref{fig:stwo} we show a neutron powder diffraction pattern collected on HB-2A for a preliminary sample which was grown using a procedure similar to S1. After our analysis of the four samples discussed in the main text we revisited this data and attempted modeling with the H intercalated structure. This produced a decent fit with nearly identical results for the crystal parameters and H content as for S1 obtained using our NOMAD data. We note that the fit here is visually somewhat worse than of those in the main text. This is a result of the different profile fitting functions - the profile used for the HB-2A data includes less modeling of crystalline disorder which as discussed is significant in these materials. 

We note that in our structural refinements of the deuterated samples (S4), in addition to the H we report on the $2b$ site, we were also able to obtain similar quality fits by including additional H on the $2a$ site (with fractional coordinates (0,0,$\frac{1}{4}$)). However, the refined occupancy was within error ($<10\%$) and its inclusion did not improve the quality of the fit as measured by the residuals. Furthermore, as will be discussed, during our Density Functional Theory analysis we found the $2a$ site to be highly energetically unfavorable. Therefore, considering this predicted unfavorable energy cost and the speciousness of including additional refinable parameters in a model without a marked improvement of the fit residuals, we do not consider reporting occupancy on this site justified and only include it here for transparency.

\newpage

\section{\label{DFT} Details of Density Functional Theory Calculations}

We performed density functional theory (DFT) calculations in order to address
the following questions:
(1) Is it energetically feasible for H to enter the structure and if
so what on what site(s)?
(2) What is the nature of the bonding of H in KCr$_3$As$_3$ and how
does this affect the electronic structure?
(3) How does this affect the lattice stability against structural
distortions? and
(4) How does H affect the proximity to magnetism?
The calculations were done with the generalized gradient approximation
(GGA) of Perdew, Burke and Ernzerhof (PBE) \cite{Perdew1996}.
We did calculations based on experimental
lattice parameters.
The energetics and phonons were determined using the
projector augmented wave (PAW) method with the VASP code.
\cite{Kresse1999,Kresse1996}
A planewave cut-off of 400 eV was used in these calculations.
We cross-checked using the general potential linearized augmented planewave
(LAPW) method as implemented in the WIEN2k code.
\cite{Singh2006,Blaha2001}
For example, the LAPW
calculated forces on the atoms were 3 mRy/Bohr or less with the relaxed
positions from VASP, and similar accord was found for other quantities.
The phonons were obtained based on PAW calculations and the
finite difference supercell method in the PHONOPY code,
\cite{Togo2015}
similar to our prior work on non-hydrided K$_2$Cr$_3$As$_3$ and
KCr$_3$As$_3$.
\cite{Taddei2018,Xing2019}
The calculations were done at a scalar relativistic level, except that
Fermi surfaces including spin-orbit are shown.

We start with the stability and site preference. We calculated energetics
for the 2$a$ (0,0,1/4); (0,0,3/4) and 2$b$ (0,0,0); (0,0,1/2)
positions, assuming full occupation of either 2$a$ or 2$b$ sites.
We find that the hydride formation enthalpy relative
to KCr$_3$As$_3$ and H$_2$ gas is positive for the 2$a$ site,
which means that it is highly unfavorable.
On the other hand a very negative formation energy of 92 kJ/mol H$_2$
is found for the 2$b$ site. This value is characteristic of a very
stable hydride that can be easily formed.
Thus both DFT and experiment are compatible with H on the 2$b$ site,
and the DFT calculations show that the formation of the hydride is
extremely favorable.
For comparison the formation
enthalpy of the stable hydride MgH$_2$ is approximately 75 kJ/mol H$_2$.

We next turn to the bonding and electronic structure. Hydrogen is an
extremely flexible chemical element, and forms common compounds as
a cation, an anion, a strongly covalent bonding, or with metallic bonding.
Here H is coordinated with Cr with relatively long bond lengths, which
would be compatible with either metallic bonding (as in PdH) or anionic
H as in MgH$_2$. In the former case, bands are distorted but H effectively
acts as an electron donor for bands near the Fermi level. In the latter
case, additional occupied H states are added, and the metallic bands
of Mg are depleted. More complicated behavior is also possible.
\cite{Gupta2005}
Cr in its bcc form does not take up significant H. However, a hexagaonal
metallic form, with stoichiometry close to CrH is readily formed
by electrochemical methods.
\cite{Khan1976}

Our electronic structure results show that the bonding is in fact
metallic. The band structure of stoichiometric
KHCr$_3$As$_3$ in comparison with non-hydrided
hexagonal KCr$_3$As$_3$ near the Fermi level $E_F$ is shown in 
the main text.
The bands are distorted for example with shifts of the bands
below the main set of bands crossing $E_F$.
However, it is clear that band filling has increased as expected for
metallic bonding of H. Thus H serves as an electron donor, and the
effective electron count of stoichiometric KHCr$_3$As$_3$ is equal to that of
K$_2$Cr$_3$As$_3$ from the point of view of the bands near $E_F$.
Thus properties simiilar to the latter compound may be anticipated.

\begin{figure}[h]
	\includegraphics[scale=0.5]{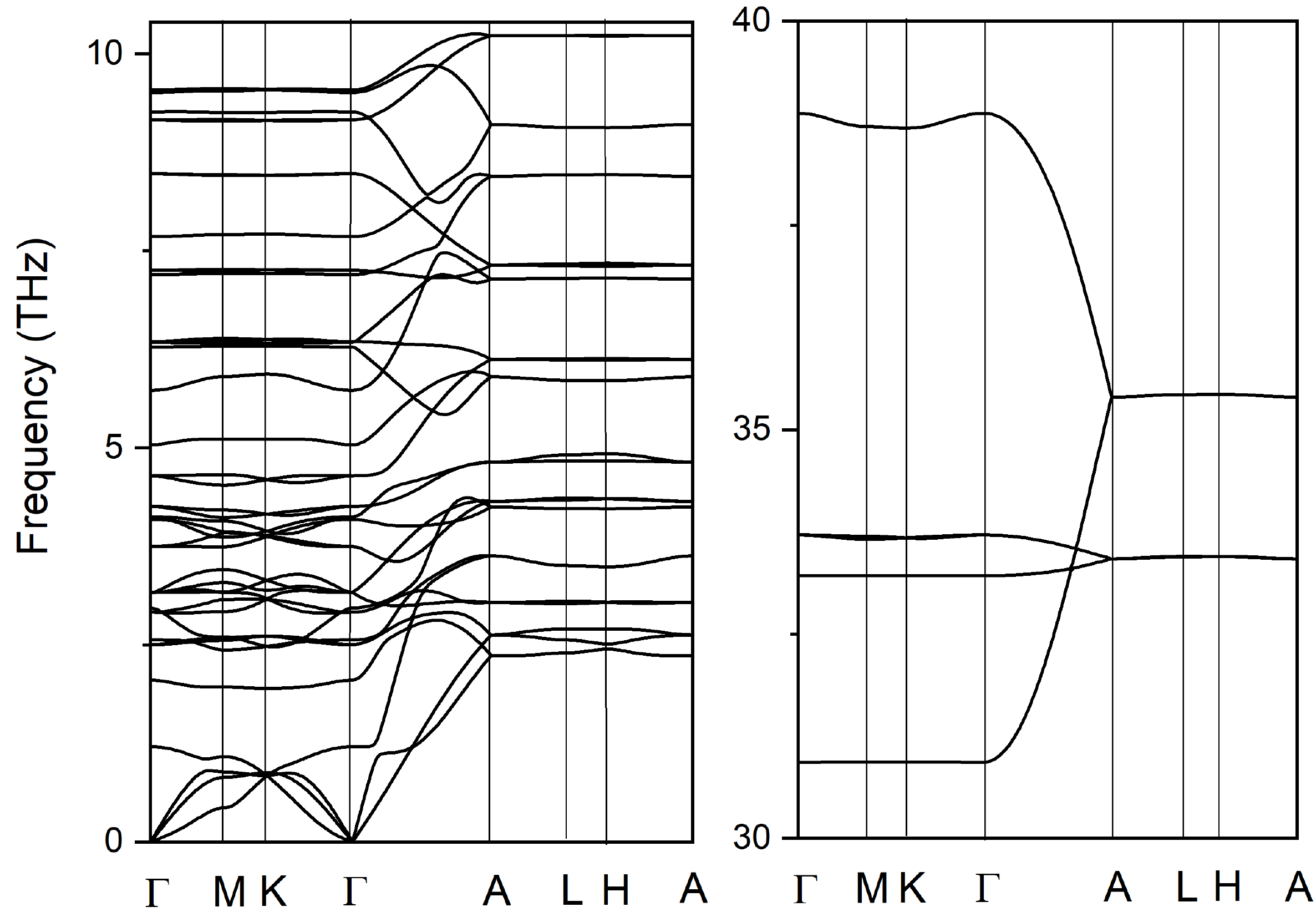}	
	\caption{\label{fig:sthree} Phonon dispersions for stiochiometric \KHoCA . }	
\end{figure} 

However, the structure is still different. In particular, the
ideal structure of stoichiometric KHCr$_3$As$_3$ is centrosymmetric,
similar to KCr$_3$As$_3$, and different from K$_2$Cr$_3$As$_3$.
This has implications for superconductivity, since singlet and triplet
channels mix in non-centrosymmetric cases but not in a centrosymmetric
structure.
In any case, K$_2$Cr$_3$As$_3$ shows structural instabilities, that
are manifested as unstable phonons in DFT calculations and 
in disordered local distortions in experiments.
\cite{Taddei2018}
Instabilities are also found in DFT calculations for KCr$_3$As$_3$.
\cite{Xing2019}
Thus phonon instabilities might be anticipated in KHCr$_3$As$_3$.
However, the instabilities in the non-hydrided compounds involve
distortion of the 1D Cr tubes. In the hydrided compound, the
H, which occurs in the center of the Cr tubes,
might modify the bonding and therefore the structural stability.
Our calculated phonons are shown in Fig. \ref{fig:sthree}.
As seen, there are no unstable modes. Therefore the H removes
the structural instability that is seen in non-hydrided K$_2$Cr$_3$As$_3$,
even though the nominal electron count is the same.
The H modes are at $\sim$35 THz, and show little dispersion except
along $\Gamma$-$A$, where
$c$-axis polarized mode disperses indicating some H-H interaction
along the tube direction. For comparison, experiments for PdD$_x$
provide an experimental D frequency of $\sim$15 THz, \cite{Rowe1974}
which would correspond to a H frequency of $\sim$21 THz.

The band structure shows three sheets of Fermi surface, which
includes a three dimensional sheet, plus two nested one dimensional
sheets. This is similar to K$_2$Cr$_3$As$_3$,
\cite{Subedi2015}
except that due to the
centrosymmetric structure, the small spin orbit splittings are absent
and all bands remain two-fold degenerate in KHCr$_3$As$_3$.
One important feature is that the three dimensional Fermi surface
sheet will make the electronic properties, including transport
distinctly three dimensional (though not isotropic) rather
than quasi-one-dimensional as might be inferred from inspection of the
crystal structure.

Finally, we did calculations to examine the proximity to magnetism.
In particular, we considered a possible ferromagnetic order as
well as an A-type order that consists of ferromagnetic layers in the Cr
wires stacked antiferromagnetically along the $c$-axis wire direction.
We find weak itinerant instability against both of these in our PBE
calculations. The A-type is lower in energy and its moments are larger.
For this state we find Cr moments of $\sim$0.3 $\mu_B$, with a magnetic
energy of $\sim$5 meV/Cr. The itinerant nature and low energy scale
suggests that quantum fluctuations could easily suppress the ordering
leading to a paramagnetic state with spin-fluctuations, analagous to
some other weak itinerant magnets, such as Ni$_3$Ga.
\cite{Aguayo2004}
In any case, the results imply that KHCr$_3$As$_3$ is near
itinerant antiferromagnetism, which could provide a pairing mechanism
in an spin-fluctuation scenario or could be a spectator in an electron-phonon
mediated scenario.

\newpage


%